%% file: main.tex
\documentclass[preprint,12pt]{elsarticle}
\journal{}
\input{packages}
\input{mycommands}

\begin{document}
\begin{frontmatter}            

\title{Non-perturbative SQED beta function using functional renormalization group approach and the NSVZ exact beta function}

\author[a,b]{Jeremy Echeverria\footnote{jeremy.echeverria.p@mail.pucv.cl}}
\author[b]{Iván Schmidt\footnote{ivan.schmidt@usm.cl}}
\affiliation[a]{organization={Instituto de Física, Pontificia Universidad Católica de Valparaíso},
            addressline={Casilla 4950}, 
            city={Valparaíso},
           country={Chile}
}
\affiliation[b]{organization={Departamento de Física, Universidad Técnica Federico Santa María y Centro Científico-Tecnológico de Valparaíso},
            addressline={Casilla 110-V}, 
            city={Valparaíso},
           country={Chile}
}

\begin{abstract}
The renormalization group equations of massive $\mathcal{N}=1$ supersymmetric quantum electrodynamics (SQED) are studied using the functional renormalization group approach. A non-perturbative form of the beta function has been computed via a derivative expansion of the effective action. In the local potential approximation, the functional form of the non-perturbative beta function is closely related to the form of the NSVZ exact beta function; this relationship is exact if an effective fine-structure constant is defined. The non-massive limit of the same is also analyzed. Furthermore, the calculation of the beta function has been improved by incorporating the influence of momentum modes on the propagation of the superfields in the non-perturbative running of the electric charge, applying a second-order truncation for the derivative expansion, which we use to find the momentum contributions to the $\gb$ function. Again, we find the NSVZ relation for an effective fine-structure constant.
\end{abstract}

\begin{keyword}
Supersymmetric quantum electrodynamics \sep Functional renormalization group \sep Exact beta function
\end{keyword}

\end{frontmatter}

\newpage

\begin{table}[!h]
    \centering
    \begin{tabular}{|c|} \hline 
        \textit{It is with sadness that I say goodbye to my professor, Iván Schmidt Andrade,}\\
        \textit{who left us during the course of this work. His passion for research and his}\\ \textit{special vision of physics work will remain with us. Thank you for everything.}\\ \hline
    \end{tabular}    
\end{table}

\newpage

\section{Introduction}
\label{sec:I}
\input{I/I.tex}

\section{$\mathcal{N}=1$ supersymmetric quantum electrodynamics}
\label{sec:sqed}
\input{1/1.tex}

\section{The functional renormalization group}
\label{sec:frg}
\input{2/2.tex}

\section{The local potential approximation}
\label{sec:lpa}
\input{3/3.tex}

\section{Second order of derivative expansion}
\label{sec:nlo}
\input{4/4.tex}

\section{Conclusions}
\label{sec:con}
\input{5/5.tex}

\section*{Acknowledgements}

We would like to thank D. Salinas-Arizmendi for his valuable advice on some calculations and P. Escalona for their useful comments and discussions. This research was partially supported by ANID PIA/APOYO AFB180002 and AFB220004 (Chile), by FONDECYT (Chile) under Grant No. 1230391. J.E thanks to ``Beca doctorado nacional'', by ANID (Chile) No. 21212381 and ``Programa de Incentivo a la Iniciación Científica'' (PIIC), by UTFSM.

\begin{appendices}

\section{Superspace formulation}
\label{sec:A}
\input{A/A.tex}

\section{Conventions, regulators and renormalized quantities}
\label{sec:B}
\input{B/B.tex}

\section{LPA' Flow}
\label{sec:C}
\input{C/C.tex}

\section{NLO Flow}
\label{sec:D}
\input{D/D.tex}

\section{NLO effective beta function and anomalous dimension}
\label{sec:E}
\input{E/E.tex}

\end{appendices}

\nocite{*}

\bibliographystyle{unsrtnat}
\bibliography{refs}

\end{document}

%% file: packages.tex
\usepackage{amssymb}
\usepackage{amsmath}
\usepackage{amsthm}
\usepackage{epsfig}
\usepackage{graphicx}
\usepackage[english]{babel}
\usepackage[utf8]{inputenc}
\usepackage{bm}
\usepackage{subcaption}
\usepackage{bbold}
\usepackage{float}
\usepackage{xcolor}
\usepackage{multirow}
\usepackage{lineno}
\usepackage[normalem]{ulem}
\usepackage{makeidx}
\usepackage{xspace}
\usepackage{wrapfig}
\usepackage{lastpage}
\usepackage{textcomp}
\usepackage{enumerate}
\usepackage{fancyhdr}
\usepackage{mathrsfs}
\usepackage{listings}
\usepackage{hyperref}
\hypersetup{backref=true,       
    pagebackref=true,               
    hyperindex=true,                
    colorlinks=true,                
    breaklinks=true,                
    urlcolor= black,                
    linkcolor= blue,                
    bookmarks=true,                 
    bookmarksopen=false,
    filecolor=black,
    citecolor=blue,
    linkbordercolor=blue
}
\usepackage{braket}
\usepackage{cancel}
\usepackage{array}
\usepackage{tabularx}
\usepackage{mathtools,slashed}
\usepackage[titletoc,title]{appendix}
\date{}

%% file: mycommands.tex
\input{greeks}

\newcommand{\al}[3][2]{
\begin{align}
    #2 \label{#3}
\end{align}}

\newcommand{\pmatx}[1]{
\begin{pmatrix}
  #1
\end{pmatrix}}

\newcommand{\nal}[1]{
\begin{align*}
    #1
\end{align*}}
\newcommand{\prt}[1]{\left( #1\right)}
\newcommand{\crt}[1]{\left[ #1\right]}
\newcommand{\lrt}[1]{\left\{ #1\right\}}

\newcommand{\prtl}[1]{\left( #1\right.}
\newcommand{\crtl}[1]{\left[ #1\right.}
\newcommand{\lrtl}[1]{\left\{ #1\right.}

\newcommand{\prtr}[1]{\left. #1\right)}
\newcommand{\crtr}[1]{\left. #1\right]}
\newcommand{\lrtr}[1]{\left. #1\right\}}
\newcommand{\artr}[1]{\left. #1\right|}
\newcommand{\dif}[3][2]{\dfrac{d #2}{d #3}}

\newcommand{\pdif}[3][2]{\dfrac{\partial #2}{\partial #3}}
\newcommand{\pddif}[3][2]{\dfrac{\partial^2 #2}{\partial #3^2}}
\newcommand{\ud}[3][2]{^{#3}_{#2}}

\newcommand{\half}{\frac{1}{2}}
\newcommand{\inv}[1]{\frac{1}{#1}}

\newcommand{\stxt}[1]{_{\textrm{#1}}}

\newcommand{\lag}{\mathcal{L}}

\newcommand{\D}{\partial}
\newcommand{\dv}[1]{\partial_{#1}}
\newcommand{\idv}[1]{\partial^{#1}}

\newcommand{\vt}[3][2]{{#2}^{#3}}
\newcommand{\ft}[3][2]{{#2}_{#3}}

\newcommand{\dd}[3][2]{\delta^{(#3)}\prt{#2}}

\newcommand{\ord}[1]{\mathcal{O}(#1)}
\newcommand{\nn}{\nonumber}

\newcommand{\hrm}[1]{{#1}^{\dagger}}

\newcommand{\td}{\Tilde}

\newcommand{\e}[1]{e^{#1}}

\newcommand{\usm}[1]{\gs^{#1}}

\newcommand{\usmb}[1]{{\bar\gs}^{#1}}

\newcommand{\lsp}[3][2]{{#2}_{#3}}
\newcommand{\lspc}[3][2]{{#2}^{#3}}
\newcommand{\rsp}[3][2]{{\bar #2}^{\dot #3}}
\newcommand{\rspc}[3][2]{{\bar #2}_{\dot #3}}
\newcommand{\cev}[1]{\reflectbox{\ensuremath{\vec{\reflectbox{\ensuremath{#1}}}}}}
\newcommand{\wdv}[1]{\tilde{\partial}_{#1}}

%% file: greeks.tex
\newcommand{\ga}{\alpha}
\newcommand{\gb}{\beta}
\newcommand{\gd}{\delta}

\newcommand{\gf}{\phi}
\newcommand{\gc}{\gamma}
\newcommand{\gh}{\eta}

\newcommand{\gj}{\xi}

\newcommand{\gl}{\lambda}
\newcommand{\gm}{\mu}
\newcommand{\gn}{\nu}
\newcommand{\go}{\theta}
\newcommand{\gp}{\pi}
\newcommand{\gq}{\psi}
\newcommand{\gr}{\rho}
\newcommand{\gs}{\sigma}
\newcommand{\gt}{\tau}

\newcommand{\gv}{\varphi}

\newcommand{\gx}{\chi}

\newcommand{\gz}{\zeta}

\newcommand{\gD}{\Delta}
\newcommand{\gF}{\Phi}
\newcommand{\gC}{\Gamma}

\newcommand{\gL}{\Lambda}

\newcommand{\gQ}{\Psi}

\newcommand{\gS}{\Sigma}

\newcommand{\gW}{\Omega}

%% file: I/I.tex
Supersymmetry is one of the leading candidates to be the fundamental theory behind the fundamental interactions of nature because, on one hand, it relates bosons and fermions through transformations that make forces and matter the same entity \cite{gol1989extension}, and on the other hand there are supersymmetric theories that unify the gauge interactions \cite{aulakh2004minimal}. Despite its astonishing theoretical achievements, no evidence for supersymmetry has been found at low energies, but there is still hope that high-energy experiments such as the Large Hadron Collider will provide clues. Some of the reasons for this hope are the elegance with which it solves theoretical problems such as the hierarchy problem \cite{martin2010supersymmetry}, the smallness of the cosmological constant \cite{froggatt2006smallness,aitchison2007supersymmetry} or the renormalizability of supergravity \cite{grisaru1976one}, as well as its essential role in string theory and the minimal supersymmetric standard model \cite{martin2010supersymmetry,aitchison2007supersymmetry,bagger1996weak}; on the other hand, the phenomenological study of supersymmetric models has shown that the particle content predicted by these models can solve current problems in particle physics; for example, through $R-$symmetry, the lightest supersymmetric particle is neutral and stable, making it an excellent candidate for dark matter \cite{jungman1996supersymmetric}.

Special cases of study are the supersymmetric extensions of the gauge theories because they are the building blocks with which supersymmetric models for the fundamental interactions beyond the standard model are built. Their supersymmetric properties give them important advantages over their non-supersymmetric counterparts, one of which is that they have a better UV behavior due to the cancellations between bosons and fermions, which gives way to non-renormalization theorems; for example, in the $\mathcal{N}=1$ supersymmetric gauge theories the superpotential is not renormalized \cite{grisaru1979improved}. Another important aspect of supersymmetric gauge theories is that they allow explicit and non-divergent condensate calculations, which are key to the non-perturbative study of vacuum states, for example when we perturb around non-trivial backgrounds such as instantons \cite{novikov1983instanton}. One feature of supersymmetric $\mathcal{N}=1$ gauge theories that follow from this is the existence of an exact beta function that relates the beta function to the anomalous dimensions of the chiral superfields, e.g. for $\mathcal{N}=1$ supersymmetric quantum electrodynamics with one flavor one has that
\al{
\gb(\ga)=\frac{\ga^2}{2\gp}\prt{1-\gc(\ga)},
}{nsvz}
where $\gc(\ga)$ is the anomalous dimension of the chiral superfield and $\ga$ the fine-structure constant. This is known as the Novikov-Shifman-Vainshtein-Zakharov or NSVZ exact beta function, named after those who first obtained it using the exact results for the instanton measure in conjunction with renormalizability \cite{novikov1983exact,novikov1986beta}. Subsequently, other researchers obtained the same exact beta function by different methods, such as the structure of anomalies \cite{jones1983more,jones1984beta,jones1984chiral,vainshtein1984axial,novikov1985supersymmetric,shifman1986solution} or the non-renormalization theorems for topological terms \cite{kraus2003supersymmetric}.

This relation has also been tested using perturbation theory in a class of renormalization schemes known as NSVZ-type schemes \cite{goriachuk2018class,goriachuk2020exact,kataev2019shell,kataev2013nsvz}, and its validity is also verified to all orders in perturbation theory, using regularization by higher derivatives \cite{stepanyantz2011derivation,kataev2013nsvz,stepanyantz2014nsvz}.

In this article we study the beta function of $\mathcal{N}=1$ supersymmetric quantum electrodynamics (SQED) using the functional (or non-perturbative) renormalization group method \cite{dupuis2021nonperturbative}, which describes the flow of the renormalization group of the effective average action (EAA) \cite{wetterich1991average,wetterich1993average,wetterich1993improvement,wetterich1993exact}, a functional that has as its infrared limit the effective action. The advantage of this approach is that it allows us to calculate the electric charge flow using a non-perturbative approximation scheme that preserves supersymmetry at all steps of the implementation and that goes beyond the perturbative approaches of the standard renormalization methods.

The application of the functional renormalization group in supersymmetric theories has been studied in some non-gauged supersymmetric models like quantum mechanics by considering the non-perturbative flow of the superpotential and the convergence of the derivative expansion of the effective average action \cite{synatschke2009flow}. Following that line, several Wess-Zumino models at zero temperature and finite temperature have been studied, in which it has been shown that the non-renormalization theorem for the superpotential is respected by the flow of the effective average action \cite{synatschke2010phase,synatschke2010two,synatschke2010n,mastaler2012supersymmetric,feldmann2016functional,feldmann2018critical}. The convergence of the derivative expansion has also been investigated for these models \cite{heilmann2015convergence}. In previous work, this method was applied to the case of a single chiral superfield coupled with a $U(1)$ vector superfield, where the behavior of the renormalization group equations was analyzed and it was found that the coupling with vector superfields does not modify the non-renormalization theorem for the superpotential \cite{echeverria2023functional}.

The main result of this study is the demonstration that the non-perturbative beta function of supersymmetric quantum electrodynamics derived from the functional renormalization group naturally coincides with the exact NSVZ beta function form of eq.(\ref{nsvz}). If the mass of the chiral superfield degrees of freedom is taken into account, i.e. if a superpotential is added, then the dimensionless mass terms modify the form of the beta function, then an NSVZ exact beta function can be constructed for an effective fine-structure constant. In the next-to-leading order case, the form of the beta function is modified by both the mass terms and the coefficients of the series expansion of the momentum modes of the field propagators, but the functional form has the form of the NSVZ.

%% file: 1/1.tex
The classical action of $\mathcal{N}=1$ supersymmetric quantum electrodynamics or SQED, including a superpotential with a mass term is
\al{
S=&\int d^4xD_{\gm}\hrm{\gf_{-}}D^{\gm}\gf_{-}+\hrm{\prt{D_{\gm}\hrm{\gf_{+}}}}D^{\gm}\hrm{\gf_{+}}+i \bar\gq_{-} \bar\gs^\gm D_{\gm}\gq_{-}+i \bar\gq_{+} \bar\gs^\gm \hrm{D_{\gm}}\gq_{+}\nn\\
&+\hrm{F_{-}}F_{-}+\hrm{F_{+}}F_{+}-\inv{4e^2}\ft{F}{\gm\gn}\vt{F}{\gm\gn}+\inv{2\gx e^2}\dv{\gm}\vt{A}{\gm}\dv{\gn}\vt{A}{\gn}+\frac{i}{e^2}\bar\gl\bar\gs^\gm\dv{\gm}\gl+\frac{1}{2e^2} D^2\nn\\
&+i\sqrt{2}\crt{\hrm{\gf_{-}}\gq_{-}\gl-\hrm{\gf}_{+}\gq_{+}\gl}+i\sqrt{2}\crt{\gf_{+}\bar\gq_{+}\bar\gl-\gf_{-}\bar\gq_{-}\bar\gl}+\prt{\hrm{\gf_{-}}\gf_{-} D-\hrm{\gf_{+}}\gf_{+}D}\nn\\
&+m\prt{\gf_-F_++\gf_+F_--\gq_-\cdot\gq_++\text{h.c}.},
}{sqed}
where the subscript ``$-$'' denotes the electron and ``$+$'' the positron, and $D_\gm=\dv{\gm}+i\ft{A}{\gm}$ is the $U(1)$ covariant derivative. This action can also be formulated in the superspace
\al{
S=&\int d^4xd^2\go d^2\bar \go\crt{\hrm{\gF_{-}}\e{-2V}\gF_{-}+\hrm{\gF_{+}}\e{2V}\gF_{+}}+\inv{32e^2}\int d^4xd^2\go W^\ga W_\ga\nn\\
&+\frac{2}{\gx e^2}\int d^4xd^2\go d^2\bar\go(D^2V)(\bar D^2V)+m\lrt{\int d^4xd^2\go\gF_-\gF_++\text{h.c}},
}{sqedsuper}
where $\gF_\mp$ are chiral superfields containing the matter fields and $V$ is a $U(1)$ vector superfield containing the gauge boson. More details in the Appendix \ref{sec:A}.

%% file: 2/2.tex
In order to obtain the non-perturbative running of the electric charge, we will compute the beta function using the functional renormalization group approach. This is constructed using the effective average action (EAA) $\gC_k$, a scale-dependent functional that interpolates between the IR and UV sectors of the theory. The main idea is that in the IR sector, the theory contains all quantum fluctuations, so for $k\to0$ the EAA tends to the ordinary effective action (EA) $\gC$, the generator of 1PI Green functions, whereas in the UV sector, characterized by a momentum cutoff $\gL$, all fluctuations are neglected, so for $k\to \gL$ the EAA will be the bare action in eq.(\ref{sqed}). The EAA will be a path in the space of differential operators with bounded conditions at $k\to0,\gL$. To respect the bound conditions, the EAA is constructed with a term,
\al{
\gD S_k[\gQ]=\int d^4q \gQ^\dagger(q)R_k(q^2)\gQ(q),
}{}
which acts as a regulator for the functional. The cutoff kernel $R_k=\gD S\ud{k}{(2)}$ is an operator containing functions that regulate the propagators of the theory and is required to be zero when $k\to0$ and infinite when $k\to\gL$. At intermediate momentum scales, $R_k$ modulates the momentum of the propagators and defines a path in the operator space for the EAA.

Starting with a bare action $S[\gv]$ and adding the regulator term, we can define the partition function as
\al{
\e{W_k[J]}=\int \mathcal{D}\gv\e{-S[\gv]-\gD S_k[\gv]-\int d^4x J(x)\gv(x)}.
}{}
To construct the EAA, we perform a Legendre transformation (for more details, see section 2.1 of \cite{gies2012introduction}),
\al{
\gC_k[\gQ]=\int d^4x J(x)\gQ(x)-W_k[J]-\gD S_k[\gQ],
}{}
where the fields $\gQ=\langle \gv\rangle_J$ are the expectation values of the fields in the presence of a current $J$ (and the reason why we can extract non-perturbative information from the theory).

The flow on the energy scale of the EAA is described by the Wetterich exact flow equation \cite{wetterich1993exact}
\al{
\dv{t}\gC_k=\half\textrm{STr}\lrt{\crt{\gC\ud{k}{(2)}+R_k}^{-1}\dv{t}R_k},
}{wetterich eq}
where $t=\ln{\frac{k}{\gL}}$ and the supertrace denotes sum over the field arrangement (see Appendix \ref{sec:B}) and momentum integration. $\gC\ud{k}{(2)}$ is the second functional derivative of the EAA with respect to the fields
\al{
\crt{\gC\ud{k}{(2)}}_{ij}(p,q)=\frac{\vec{\gd}}{\gd\gQ_i(p)}\gC_k\frac{\cev{\gd}}{\gd\gQ_j(q)}.
}{}
To ensure that EAA preserves supersymmetry in all steps of its implementation, we need to add a regulating term that respects supersymmetry. According to \cite{echeverria2023functional}, for a supersymmetric gauge theory, we can use the regulating term
\al{
\gD S_k=&\int d^4xd^2\go d^2\bar \go\crt{\hrm{\gF_{-}}r(\square)\e{-2V}\gF_{-}+\hrm{\gF_{+}}r(\square)\e{2V}\gF_{+}}\nn\\
&+\inv{32}\int d^2\go\vt{W}{\ga} t(\square)\ft{W}{\ga},
}{regact}
where the functions $r$ and $t$ are the optimized Litim-type regulator functions \cite{litim2000optimisation} for fermions and bosons. The explicit forms of the dimensionless and renormalized regulators (more details in the Appendix \ref{sec:B}) in momentum space are
\al{
r=&\prt{\dfrac{1}{q}-1}\go(1-q^2)\nn\\
t=&\prt{\dfrac{1}{q^2}-1}\go(1-q^2).
}{reg}
In order to extract useful information from the EAA flow, it is possible to use several types of truncation schemes, in which the dependence on one parameter of the theory is truncated while the full dependence on another parameter is retained (a more detailed explanation can be found in the introduction of \cite{baldazzi2021functional}). Since we are interested in obtaining a non-perturbative beta function, we will use a truncation of the derivative expansion of the EAA, which allows us to preserve the dependence on the coupling of the beta function at all orders of the power expansion at the cost of truncating the dependence on the high-energy momentum modes. However, in order to get a first look at how such momentum modes modify the flow of the coupling constant, we will study the next-to-leading order of the expansion.

%% file: 3/3.tex
In the leading order of the derivative expansion, only the superpotential is considered to be scale dependent, this framework is known as ``local potential approximation'' (LPA). In this case, this truncation is insufficient because, as we will see later, the beta function depends on the anomalous dimension of the vector superfield, so following the procedure of algebraic renormalization, the fields, and electric charge have been re-escalated using wave function renormalization functions that are independent of both the fields and the energy scale,
\nal{
\gF_0\to {Z_\gF}^{\half} \gF,\quad V_0\to {Z_V}^{\half} V,\quad e_0\to Z_e e.
}

This improvement of LPA is called LPA' truncation and is our leading order in the supercovariant derivative expansion.

To avoid violating gauge symmetry, the re-scaling function of the electric charge cannot be independent of the other functions, in particular, $Z_e={Z_V}^{-\frac{1}{2}}$, which gives us the following relation
\al{
e_0=Z_e e={Z_V}^{-\frac{1}{2}} e.
}{}
As a consequence, the beta function is defined as
\al{
\gb(e)=\dif{e}{\ln{\frac{k}{\gL}}}=-\half\gh_{V}e
}{}
in $d=4$. The term $\gh_\mathcal{O}=-\dv{t}\ln{Z_\mathcal{O}}$ is the anomalous dimension of the operator $\mathcal{O}$. Similarly, the anomalous dimension of the chiral superfield is given by
\al{
\gc=\gh_\gF=-\dif{\ln Z_\gF}{\ln{\frac{k}{\gL}}}.
}{canoma}
The LPA' truncation for the effective action has the form
\al{
\gC_k=&Z_\gF\int d^4xd^2\go d^2\bar \go\crt{\hrm{\gF_{-}}\e{-2V}\gF_{-}+\hrm{\gF_{+}}\e{2V}\gF_{+}}+\frac{{Z_V}^2}{32e^2}\int d^4xd^2\go W^\ga W_\ga\nn\\
&+\frac{2{Z_V}^2}{\gx e^2}\int d^4xd^2\go d^2\bar\go(D^2V)(\bar D^2V)+m_0\lrt{\int d^4xd^2\go\gF_-\gF_++\text{h.c}},
}{}
or, in components of superfields
\al{
\gC_k=&\int d^4xZ_{\gF}\crtl{D_{\gm}\hrm{\gf_{-}}D^{\gm}\gf_{-}+\hrm{\prt{D_{\gm}\hrm{\gf_{+}}}}D^{\gm}\hrm{\gf_{+}}+i \bar\gq_{-} \bar\gs^\gm D_{\gm}\gq_{-}+i \bar\gq_{+} \bar\gs^\gm \hrm{D_{\gm}}\gq_{+}}\nn\\
&\crtr{+\hrm{F_{-}}F_{-}+\hrm{F_{+}}F_{+}}+\frac{{Z_{V}}^2}{e^2}\crt{-\inv{4}\ft{F}{\gm\gn}\vt{F}{\gm\gn}+\inv{2\gx e^2}\dv{\gm}\vt{A}{\gm}\dv{\gn}\vt{A}{\gn}+i\bar\gl\bar\gs^\gm\dv{\gm}\gl+\frac{1}{2} D^2}\nn\\
&+i\sqrt{2}Z_{\gF}\lrt{\crt{\hrm{\gf_{-}}\gq_{-}\gl-\hrm{\gf}_{+}\gq_{+}\gl}+i\sqrt{2}\crt{\gf_{+}\bar\gq_{+}\bar\gl-\gf_{-}\bar\gq_{-}\bar\gl}}\nn\\
&+Z_{\gF}\prt{\hrm{\gf_{-}}\gf_{-} D-\hrm{\gf_{+}}\gf_{+}D}+m_0\prt{\gf_-F_++\gf_+F_--\gq_-\cdot\gq_++\text{h.c}.}.
}{lpa}
The LPA' truncation of the coupling between chiral and vector superfields respects the non-renormalization theorem for the superpotential \cite{echeverria2023functional}, so we can simply truncate the superpotential without losing any information; on the other hand, the wave function renormalization function for the electron and positron are the same, because both are components of a Dirac fermion.

Since the renormalization functions of the wave function have no dependence on the energy scale, we can restrict all fields to real constants in momentum space $\gQ_i(q)=(\sqrt{2\gp})^4\gQ_i\gd(q)$ without affecting the flow of the renormalization functions. To extract the corresponding information, we have to project eq.(\ref{wetterich eq}) in a suitable configuration of the fields in such a way that it is possible to isolate the term containing the information. For example, projecting on $\gf_-=\gf_+=\gq_-=\gq_+=F_+=F_-=A=\gl=0$ we get
\al{
\dv{t}\gC_k=(\sqrt{2\gp})^4\dd{0}{4}\dv{t}\crt{\frac{{Z_{V}}^2}{2e^2}D^2}.
}{}
From this point, we can see that
\al{
-\frac{1}{2}\gh_V=\frac{e^2}{4{Z_V}^2}\artr{\lrt{\artr{\pddif{}{D}\dv{t}\gC_k}_{\gf_-=\gf_+=\gq_-=\gq_+=F_+=F_-=A=\gl=0}}}_{D=0}.
}{betaeqlpa}
In the same way we can project eq.(\ref{lpa}) onto $F_-=F_+=\gq_-=\gq_+=A=\gl=0$ in order to get,
\al{
\dv{t}\gC_k=(\sqrt{2\gp})^4\dd{0}{4}\dv{t}\crt{Z_{\gF}\hrm{\gf_-}\gf_-D+\frac{{Z_{V}}^2}{2e^2}D^2},
}{}
therefore\footnote{Note that in eq.(\ref{betaeqlpa}) and eq.(\ref{anomalpa}) we are also taking into consideration the conjugate form of the fields.},
\al{
\gc=-\frac{1}{{Z_\gF}}\artr{\lrt{\artr{\frac{\D^3}{\D\gf_-\D\hrm{\gf_-}\D D}\dv{t}\gC_k}_{F_-=F_+=\gq_-=\gq_+=A=\gl=0}}}_{\gf_-=D=0}.
}{anomalpa}
The right-hand sides of eq.(\ref{betaeqlpa}) and eq.(\ref{anomalpa}) have been obtained from eq.(\ref{wetterich eq}) in momentum space
\al{
\dv{t}\gC_k=\frac{1}{2}\int d^4q \textrm{Tr}\lrt{\prt{\crt{\gC\ud{k}{(2)}}(p,q)+R_k(p,q)}^{-1}\dv{t}R_k(p,q)\dd{p-q}{4}}.
}{wetterichlpa}
Since all the fields are constant, we can simply invert the matrix in Appendix \ref{sec:C} and take the matrix trace. For the momentum integration, both the regulator functions as well as the energy scale, mass and the coupling constant have been expressed in a dimensionless and renormalized way (see Appendix \ref{sec:B}). This leads to the anomalous dimension of the chiral superfield
\al{
\gc(\ga)=\frac{5 \alpha\left(\ga(1+m^2)-6\pi\left(1-m^2\right)^2\right)}{5\left(1+m^2\right) \alpha^2-9\pi\alpha\left(1-m^2\right)^2\left(1+m^2\right)+60\pi^2\left(1-m^2\right)^5}
}{lpagamma}
and the flow equation for the coupling
\al{
\gb(\ga)=\frac{3 \alpha^2\left(\left(7+4 m^2-3 m^4\right) \alpha+20\left(1-m^2\right)^3\left(1+m^2\right) \pi\right)}{10\alpha^2\left(1-m^4\right)-18\gp\ga\left(1-m^2\right)^3\left(1+m^2\right)+120\pi^2\left(1-m^2\right)^6}
}{lpabeta}
where $\ga$ is the renormalized fine-structure constant $\ga=e^2/4\gp$. The main result of this work is that eq.(\ref{lpabeta}) can be writing in the form
\al{
\gb(\ga)=\frac{\ga^2}{2\gp}\frac{(1+m^2)}{(1-m^2)^3}\prt{1-\gc(\ga)},
}{lpansvz}
which is a modification of the NSVZ relation in eq.(\ref{nsvz}) by a dimensionless mass term.

\subsection{Fixed points}
The study of fixed points, i.e. values of the fine-structure constant $\ga_*$ for which $\gb(\ga_*)=0$, is fundamental. If the beta function possesses stable nontrivial fixed points, then there exist certain configurations of the parameters that yield a well-defined theory around it; if, in addition, there exists a path connecting the gaussian (trivial) fixed point with the non-trivial one, then the theory can be said to have asymptotic safety \cite{percacci2009asymptotic} and well defined at high energies.

The exact form of the beta function in eq.(\ref{lpansvz}) shows us that fixed points can be found equivalently as those for which $\gc(\ga_*)=1$. Clearly, there is a trivial fixed point for $\ga_*=0$, and with respect to non-trivial fixed points, it can be seen that eq.(\ref{lpabeta}) vanishes for
\al{
\ga_*=-\frac{20\pi\left(-1+3 m^2-3 m^4+m^6\right)}{-7+3 m^2},
}{fixedpoints}
or also $\gc(\ga_*)=1$ for these values. Since $\ga$ depends on the square of the electric charge, we are only interested in fixed points that represent $\ga_*>0$. 

From eq.(\ref{fixedpoints}) we can see that there are values of the square of the mass for which $\ga$ is positive, and that there is also a value of the mass that causes a discontinuity. Numerically, the range in which there will be valid non-trivial fixed points is $m\in (1,1.5275')$.

\begin{figure}[h]
    \centering
    \includegraphics[width=1\linewidth]{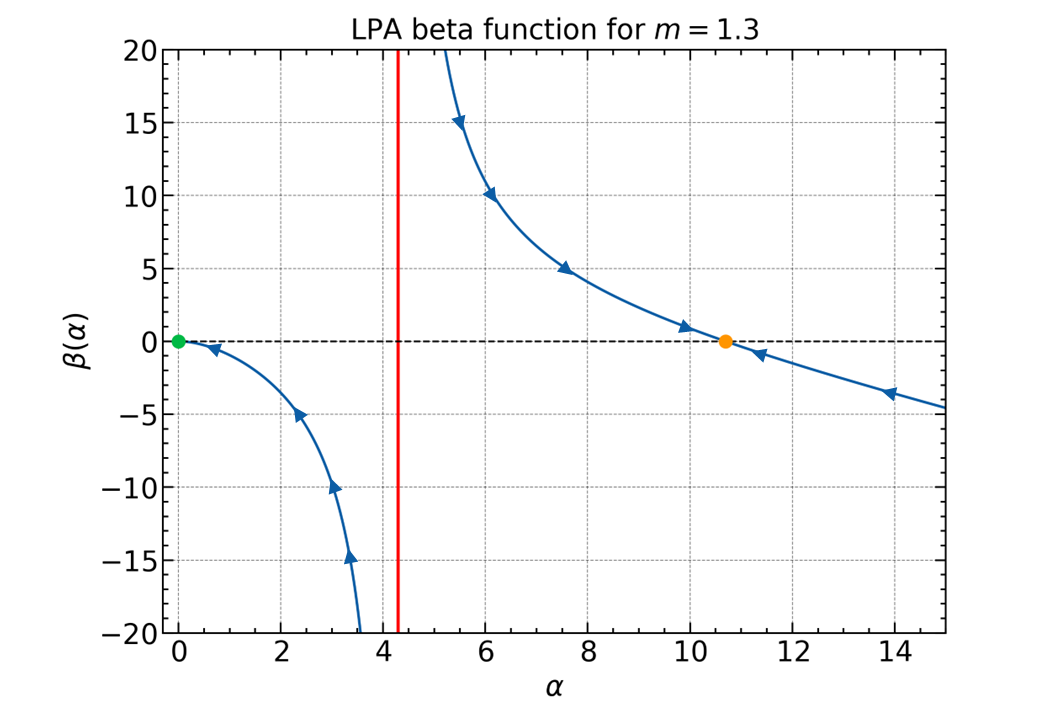}
    \caption{Flow diagram of the beta function in eq.(\ref{lpabeta}) for $m=1.3$. The green dot corresponds to the Gaussian fixed point and the orange dot corresponds to the non-trivial fixed point at $\ga_*=10.6947$. The discontinuity of the beta function for $\ga=4.2954'$ is represented by the red vertical line.}
    \label{fig:betafixedpoints}
\end{figure}

In Fig. \ref{fig:betafixedpoints} we can see the form of the beta function for a value of the mass in the range of validity, in it we can clearly identify two regions separated by an asymptote, where in one of them lives the trivial fixed point and in the other a non-trivial fixed point, the non-trivial fixed point is stable in that small perturbations around it bring us back to the fixed point, however there are no trajectories connecting both points, therefore we cannot say that this theory possesses configurations with asymptotic safety.

\subsection{The non-massive limit}
\begin{figure}[h]
    \centering
    \includegraphics[width=0.8\linewidth]{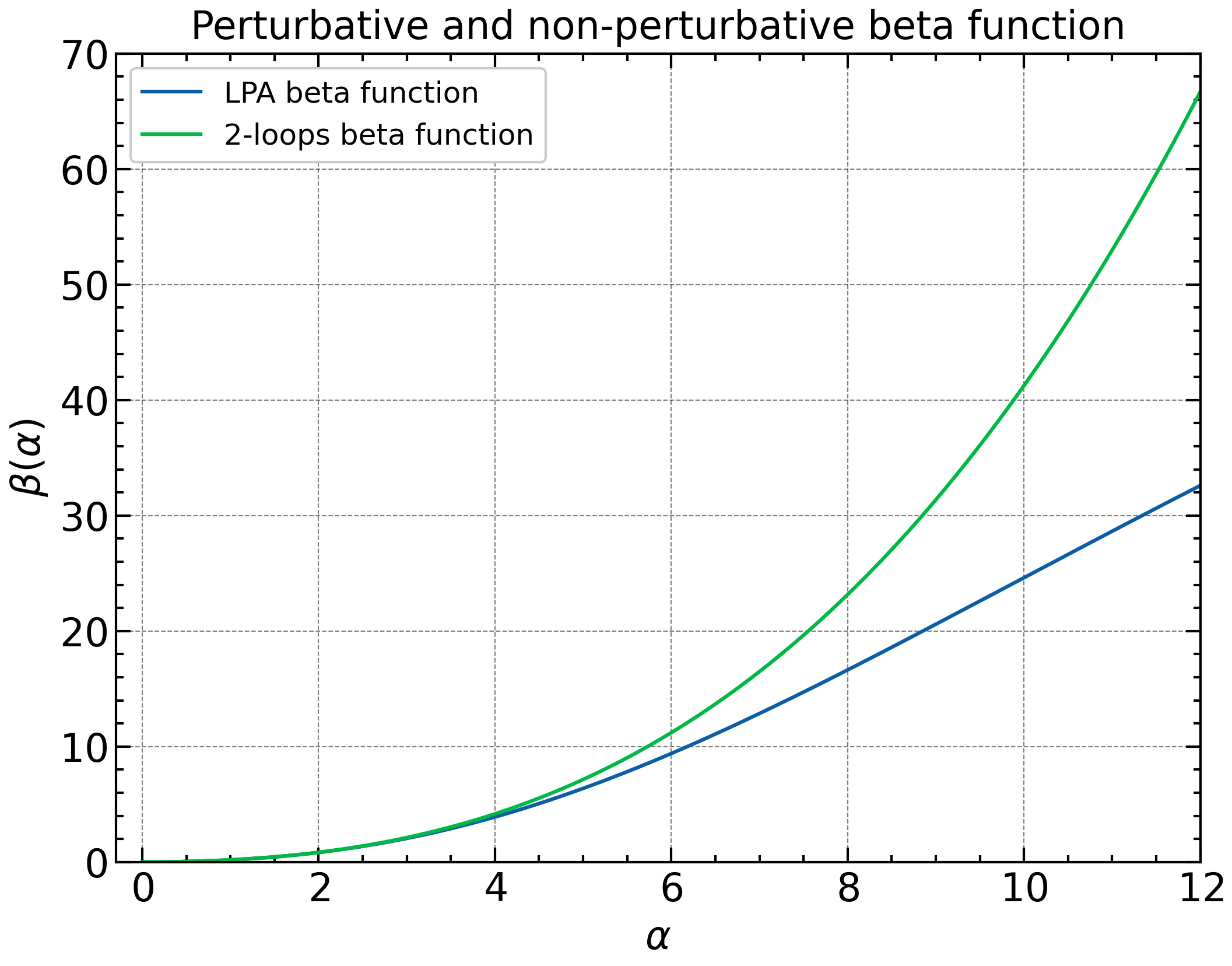}
    \caption{Comparison between the beta function in LPA and the 2-loops beta function.}
    \label{fig:lpavs2l}
\end{figure}

The non-massive SQED limit corresponds to switching off the superpotential, or in other words to setting $m=0$ in eq.(\ref{lpagamma}), eq.(\ref{lpabeta}) and eq.(\ref{lpansvz}) which leaves us with the following result
\al{
\gc(\ga)=&\frac{5\alpha(\alpha-6\pi)}{5\alpha^2-9\alpha\pi+60 \pi^2},\nn\\
\gb(\ga)=&\frac{3 \alpha^2(7 \alpha+20 \pi)}{10\alpha^2-18\alpha\pi+120\pi^2}.
}{gblpa}
The beta function is consistent with the perturbative result in e.g. the two-loop beta function of eq.(2.2.81) in \cite{seijas2007beta} (in powers of the electric charge)
\al{
\gb_e({e})=\frac{{e}^3}{4(4\pi^2)}+\frac{{e}^5}{8(4\pi^2)^2}+\ord{{e}^7}.
}{}
Futhermore, they satisfy the NSVZ relation,
\al{
\gb(\ga)=\frac{\ga^2}{2\gp}\prt{1-\gc(\ga)},
}{nsvzsnm}
which can be seen expanding eq.(\ref{gblpa}) in powers of $\ga$,
\al{
\gb(\ga)=&\frac{\ga^2}{2\gp}+\frac{\ga^3}{4\gp^2}-\frac{\ga^4}{240\gp^3}+\ord{\ga^5}\nn\\
\gc(\ga)=&-\frac{\ga}{2\gp}+\frac{\ga^2}{120\gp^2}+\ord{\ga^3}.
}{}
From eq.(\ref{fixedpoints}) we can see that there are no fixed points of the beta function for this limit. A comparison between the perturbative beta function at 2-loops and the non-perturbative one in LPA' can be seen in Fig.\ref{fig:lpavs2l}.

\subsection{The massive case and the beta function in terms of an effective coupling constant}
Keeping the superpotential, we get contributions from the dimensionless mass terms in the beta function, as in eq.(\ref{lpansvz}). However, it is possible to modify it to obtain the functional form of a NSVZ beta function by defining the following effective constant
\al{
\td\ga^2={\ga}^2\frac{(1+m^2)^2}{(1-m^2)^6}
}{}
where $\ga$ is the renormalized fine-structure constant. Redefining the beta function and the anomalous dimension in the form
\al{
\td\gc(\td\ga)&=\gc(\td\ga)\nn\\
\td\gb(\td\ga)&=\frac{(1+m^2)}{(1-m^2)}\gb(\td\ga),
}{}
we obtain
\al{
&\td\gc(\td\ga)=\frac{5 \td\alpha\left(\td\ga(1+m^2)-6\pi\right)}{5\left(1-m^2\right) \td\alpha^2-9\pi\td\alpha\left(1+m^2\right)^2\left(1+m^2\right)+60\pi^2\left(1+m^2\right)}\nn\\
&\td\gb(\td\ga)=\frac{3 \td\alpha^2\left(\left(7-3m^2\right) \td\alpha+20\left(1+m^2\right) \pi\right)}{10\td\alpha^2\left(1-m^2\right)+18\gp\td\ga\left(1+m^2\right)-120\pi^2\left(1+m^2\right)}.
}{}
In the non-massive limit they are equivalent to eq.(\ref{gblpa}) and satisfy eq.(\ref{nsvz})
\al{
\td\gb(\td\ga)=\frac{\td\ga^2}{2\gp}\prt{1-\td\gc(\td\ga)},
}{}
which can be seen expanding in powers of $\td\ga$
\al{
&\td\gb(\td\ga)=\frac{\td\ga^2}{2\gp}+\frac{1}{2(m^2+1)}\frac{\td\ga^3}{2\gp^2}+\frac{(1-10m^2)}{2(m^2+1)}\frac{\td\ga^4}{120\gp^3}+\ord{\td\ga^5}\nn\\
&\td\gc(\td\ga)=-\frac{1}{(m^2+1)}\frac{\td\ga}{2\gp}-\frac{(1-10m^2)}{(m^2+1)}\frac{\td\ga^2}{120\gp^2}+\ord{\td\ga^3}.
}{}

%% file: 4/4.tex
In order to improve the calculation of the beta function by adding the effects of momentum modes of the order of the wave functions we will use the second order in the derivative expansion, using arbitrary functions $Z_k(\square)$ modulating the kinetic part of the Lagrangian, which allows us to add the influence of such momentum modes in the propagation of both superfields in the running of the electric charge.

The second-order truncation for the effective action will be
\al{
\gC_k=&\int d^4xd^2\go d^2\bar \go Z_{\gF,k}(\square)\crt{\hrm{\gF_{-}}\e{-2V}\gF_{-}+\hrm{\gF_{+}}\e{2V}\gF_{+}}\nn\\
&+\frac{1}{32e^2}\int d^4xd^2\go {Z_{V,k}(\square)}^2W^\ga W_\ga+\frac{2{Z_{V,k}(\square)}^2}{\gx e^2}\int d^4xd^2\go d^2\bar\go(D^2V)(\bar D^2V)\nn\\
&+m_0\lrt{\int d^4xd^2\go\gF_-\gF_++\text{h.c}}.
}{}
To respect the results of the local potential approximation, we require that the arbitrary functions are the ordinary wave function renormalization functions in the limit $q\to0$ (in momentum space),
\al{
Z_{\gF/V,k}(q\to0)\to Z_{\gF/V}
}{}
which has been implemented by parameterizing the momentum information in a function $\gz_{\gF/V,k}(q)$ with momentum dependence \cite{feldmann2018critical} such that
\al{
Z_{\gF/V,k}(q)=Z_{\gF/V}\gz_{\gF/V,k}(q)
}{}
with $\gz_{\gF/V,k}(0)=1$.

Since this time the renormalization functions of the wave function have a dependence on the momentum, we cannot restrict the totality of the field to real constants; for example, if we project the EAA in the term proportional to $D^2$ and we make $D(q)$ a real constant, this will happen
\al{
\inv{(\sqrt{2\gp})^4}\int d^4 q \frac{{Z_{V,k}(q)}^2}{2e^2}D(q)D(-q)=\dd{0}{4}\frac{{Z_{V}}^2}{2e^2}\gz_{V,k}(0)^2D^2
}{}
and we have lost all information about the momentum modes, so we will restrict all fields except $D(q)$. Projecting eq.(\ref{wetterich eq}) on $\gf_-=\gf_+=\gq_-=\gq_+=F_+=F_-=A=\gl=0$ we obtain
\al{
\dv{t}\gC_k=\int d^4 p \dv{t}\crt{\frac{{Z_{V,k}(p)}^2}{2e^2}D(p)D(-p)}.
}{}
From this point, we can see that
\al{
\frac{1}{2Z_V}&\dv{t}Z_{V,k}(p)=-\half \gh_V\gz_{V,k}(p)+\half\dv{t}\gz_{V,k}(p)\nn\\
=&\frac{e^2}{4Z_VZ_{V,k}(p)}\artr{\lrt{\artr{\frac{\gd^2}{\gd D(p)\gd D(-p)}\dv{t}\gC_k}_{\gf_-=\gf_+=\gq_-=\gq_+=F_+=F_-=A=\gl=0}}}_{D=0}.
}{betanlo}
In the same way we can project onto $F_-=F_+=\gq_-=\gq_+=A=\gl=0$, keeping $\gF_-(q)$, $\hrm{\gF_-}(q)$ and $D(q)$ as momentum-dependent and we get
\al{
\dv{t}\gC_k=\int d^4 p \dv{t}\crt{\int d^4p'{Z_{\gF,k}(p)}\gF_-(p)\hrm{\gF_-}(p')D(p'-p)+\frac{{Z_{V,k}(p)}^2}{2e^2}D(p)D(-p)},
}{}
therefore,
\al{
\frac{1}{Z_\gF}&\dv{t}Z_{\gF,k}(p)=-\gc\gz_{\gF,k}(p)+\dv{t}\gz_{\gF,k}(p)\nn\\
=&\frac{1}{Z_\gF}\artr{\lrt{\artr{\frac{\gd^2}{\gd \gF_-(p)\gd \hrm{\gF_-}(p')\gd D(p'-p)}\dv{t}\gC_k}_{F_-=F_+=\gq_-=\gq_+=A=\gl=0}}}_{\gf_-=D=0}.
}{anomanlo}
The right-hand sides of eq.(\ref{betanlo}) and eq.(\ref{anomanlo}) have been obtained from eq.(\ref{wetterich eq}) in momentum space
\al{
\dv{t}\gC_k=\frac{1}{2}\int d^4q \textrm{Tr}\lrt{\prt{\crt{\gC\ud{k}{(2)}}(p,q)+R_k(p,q)}^{-1}\dv{t}R_k(p,q)}
}{wetterichnlo}
with the important difference that in this case, the operators have non-trivial dependence on momentum, and therefore we cannot compute eq.(\ref{wetterichnlo}) taking the inverse matrix of the operator $G_k=\gC_k^{(2)}+R_k$. The calculation strategy is to decompose the regularized inverse propagator operator $G_k$ in a field-independent part $G_{0,k}$ and in a field-dependent part $\gD G_{k}$, as is shown in \cite{gies2002renormalization}
\al{
G_k=G_{0,k}+\gD G_{k},
}{desc}
whose specific forms are given in the Appendix \ref{sec:D}. Substituting this into eq.(\ref{wetterich eq}) allows us to perform an expansion of the Wetterich equation in the field-dependent part of the operator
\al{
\dv{t}\gC_k=&\half\text{STr}\wdv{t}\ln{\crt{G_{0,k}+\gD G_{k}}}\nn\\
=&\half\text{STr}\wdv{t}\prt{{G_{0,k}}^{-1}\gD G_{k}}-\inv{4}\text{STr}\wdv{t}\prt{{G_{0,k}}^{-1}\gD G_{k}}^2\nn\\
&+\inv{6}\text{STr}\wdv{t}\prt{{G_{0,k}}^{-1}\gD G_{k}}^3+\ldots
}{wetterichnloex}
where $\wdv{t}$ only affects the operator $R_k$. The inverse of the field-independent part can be computed by simply matrix inversion. Since eq.(\ref{betanlo}) is quadratic and eq.(\ref{anomanlo}) is cubic in the fields, we only need to expand up to the third order in the field-dependent part. After taking the matrix trace and performing the functional derivatives, we obtain the flow of the momentum-dependent wave function renormalization functions before the momentum integration (more details in Appendix \ref{sec:D})
\al{
&\frac{1}{2Z_V}\dv{t}Z_{V,k}(p)=-\frac{\gW_d}{(2\gp)^d}\int\ud{0}{1} dqq^{d-1}\frac{e^2\gr(q-p)\mathcal{M_+}(q)\dv{t}r(q)}{\mathcal{M_-}(q)^2\mathcal{M_-}(q-p))}\nn\\
&\frac{1}{Z_\gF}\dv{t}Z_{\gF,k}(p)=\frac{\gW_d}{(2\gp)^d}\int\ud{0}{1} dqq^{d-1}\prtl{e^2\crtl{\dv{t}t(q)\gt(q-p)\gr(q-p)\mathcal{M_-}(q)^2(\gr(q-p)+2\gr(q))}}\nn\\
&-2\dv{t}r(q)\gt(q)\mathcal{M_+}(q)(\gt(q)\gr(q-p)\mathcal{M_-}(q-p)+\gt(q-p)\gr(q-p)\mathcal{M_-}(q)\nn\\
&\prtr{\crtr{+\gt(q)\gr(q)\mathcal{M_-}(q-p))}}\times \prt{3\gt(q)^2\gt(q-p)\mathcal{M_-}(q)^2\mathcal{M_-}(q-p)^2}^{-1},
}{nloint}
where
\al{
&\gr(q)=r(q)+1\nn\\
&\gt(q)=t(q)+1\nn\\
&\mathcal{M}_\pm(q)=m_0^2\pm q^2\gr(q)^2.
}{}
At this point the functions $\gz_{\gF/V,k}(p)$ remain arbitrary. To extract explicit information about the momentum modes we will polynomially truncate $\gz_{\gF/V,k}(p)$ \cite{feldmann2016functional}
\al{
\gz_{\gF/V,k}(p)=\sum\ud{n=0}{\infty}\gz_{\gF/V,n}p^n
}{z}
with $\gz_{\gF/V,0}=1$. The beta function and the anomalous dimension are obtained replacing $p=0$ in eq.(\ref{nloint})
\al{
&\frac{1}{2Z_V}\dv{t}Z_{V,k}(0)=-\half \gh_V\gz_{V,k}(0)+\half\dv{t}\gz_{V,k}(0)=-\half \gh_V=\inv{e}\gb(e)\nn\\
&\frac{1}{Z_\gF}\dv{t}Z_{\gF,k}(0)=-\gc\gz_{\gF,k}(0)+\dv{t}\gz_{\gF,k}(0)=-\gc.
}{}
For example, we will see what happens in the linear case, $\gz_{\gF/V}(p)=1+\gz_{\gF/V,1}p$. Performing $p=0$ in eq.(\ref{nloint}) and solving the integrals we get
\al{
\gc=&\frac{g_{0,2}\ga^2+g_{0,1}\ga}{g_{1,2}\ga^2+g_{1,1}\ga+g_{1,0}}
}{nlogamma}
and
\al{
\gb(\ga)=&\frac{b_{0,3}\ga^3+b_{0,2}\ga^2}{b_{1,2}\ga^2+b_{1,1}\ga+b_{1,0}}
}{}
where
\al{
g_{0,2}&=-5(m^2+(1+\gz_{\gF,1})^2)(1+\gz_{\gF,1})^3(2+3\gz_{\gF,1})(3\gz_{V,1}(7\gz_{V,1}+12)+14)\nn\\
g_{0,1}&=10\gp(1+\gz_{\gF,1})(m^2+(1+\gz_{\gF,1})^2)^2\crtl{8\gz_{V,1}(m^2(-3+4\gz_{\gF,1})+(1+\gz_{\gF,1})^2(24+31\gz_{\gF,1}))}\nn\\
&+7{\gz_{V,1}}^2(m^2(-3+\gz_{\gF,1})+(1+\gz_{\gF,1})^2(15+19\gz_{\gF,1}))\nn\\
&\crtr{+28\gz_{\gF,1}(m^2+\gz_{\gF,1}(11+4\gz_{\gF,1})+10)+84}\nn\\
g_{1,2}&=g_{0,2}\nn\\
g_{1,1}&=84\gp(m^2-(1+\gz_{\gF,1})^2)(m^2+(1+\gz_{\gF,1})^2)(1+\gz_{\gF,1})(3+5\gz_{\gF,1})(1+\gz_{V,1})^2\nn\\
g_{1,0}&=1680\gp^2(m^2-(1+\gz_{\gF,1})^2)^5(1+\gz_{V,1})^4,\nn\\
b_{0,3}&=(1+\gz_{\gF,1})^2(2+3\gz_{\gF,1})(m^2+(1+\gz_{\gF,1})^2)\lrtl{(1+\gz_{\gF,1})^2\crtl{\gz_{V,1}(7\gz_{V,1}(57+65\gz_{\gF,1})}}\nn\\
&+820\gz_{\gF,1}+708)\crtr{+350\gz_{\gF,1}+294}-m^2\crtl{5\gz_{V,1}\gz_{\gF,1}(52+35\gz_{V,1})}\nn\\
&\lrtr{\crtr{+3\gz_{V,1}(124+77\gz_{V,1})+70\gz_{\gF,1}+126}}\nn\\
b_{0,2}&=-840\gp(1+\gz_{V,1})^4(1+\gz_{\gF,1})^2(2+3\gz_{\gF,1})(m^2-(1+\gz_{\gF,1})^2)(m^2+(1+\gz_{\gF,1})^2)\nn\\
b_{1,2}&=-10(1+\gz_{\gF,1})^3(2+3\gz_{\gF,1})(14+13\gz_{V,1}(12+7\gz_{V,1}))\nn\\
&\times(m^2-(1+\gz_{\gF,1})^2)(m^2+(1+\gz_{\gF,1})^2)\nn\\
b_{1,1}&=168\gp(1+\gz_{V,1})^2(1+\gz_{\gF,1})(3+5\gz_{\gF,1})(m^2-(1+\gz_{\gF,1})^2)(m^2+(1+\gz_{\gF,1})^2)\nn\\
b_{1,0}&=3360\gp^2(m^2-(1+\gz_{\gF,1})^2)^6(1+\gz_{V,1})^4.
}{}
The beta function can be expressed in an NSVZ form
\al{
\gb(\ga)=-\frac{\ga^2}{2\gp}\prt{\frac{(m^2+(1+\gz_{\gF,1})^2)}{2(m^2-(1+\gz_{\gF,1})^2)^3}}(1+\gz_{\gF,1})(2+3\gz_{\gF,1})\prt{1-\gc(\ga)},
}{nlonsvz}
which reduces to the expresion eq.(\ref{lpansvz}) when $\gz_{\gF,1}\to0$. Here we see that the next-to-leading order adds the influence of the momentum modes on the beta function; however, its form is still of the NSVZ relation type.

We will have, in addition to the mass, contributions from the momentum modes, but we can follow the same procedure and define an effective fine-structure constant. In this case, the effective constant will be
\al{
\td\ga^2=-{\ga}^2\prt{\frac{(m^2+(1+\gz_{\gF,1})^2)}{2(m^2-(1+\gz_{\gF,1})^2)^3}}(1+\gz_{\gF,1})(2+3\gz_{\gF,1})
}{effcte}
Redefining the beta function and the anomalous dimension in the form
\al{
\td\gc(\td\ga)&=\gc(\td\ga)\nn\\
\td\gb(\td\ga)&=-\prt{\frac{(m^2+(1+\gz_{\gF,1})^2)}{2(m^2-(1+\gz_{\gF,1})^2)^3}}(1+\gz_{\gF,1})(2+3\gz_{\gF,1})\gb(\td\ga),
}{effgbm}
we can see that this respects the form of eq.(\ref{nsvz})
\al{
\td\gb(\td\ga)=\frac{\td\ga^2}{2\gp}\prt{1-\td\gc(\td\ga)},
}{}
where the explicit form of eq.(\ref{effgbm}) is given in Appendix \ref{sec:E}. An important point to note is that this procedure can be done for a polynomial truncation of arbitrary order in eq.(\ref{z}), therefore, it is always possible to enforce the relation. This relation is a consequence of the supersymmetric structure of the vacuum of the theory, so that the structure of the momentum modes can still be represented in this relation even when we increase the truncation level is a direct consequence of the fact that the method preserves supersymmetry in its implementation.

%% file: 5/5.tex
In this work, the beta function of $\mathcal{N}=1$ supersymmetric quantum electrodynamics has been calculated in the leading and next-to-leading order of the supercovariant derivative expansion of the effective average action, in order to study its non-perturbative behavior. 

In the leading order, the local potential approximation, we truncated the EAA by assuming that only the superpotential has an energy dependence, and we also considered the wave function renormalization functions to access the anomalous dimensions of the fields. Here we obtained a relation between the beta function and the anomalous dimension as in the NSVZ relation, eq.(\ref{lpansvz}), with the difference of a dimensionless mass-dependent factor, which motivated us to define an effective fine-structure constant whose beta function respects that relation.

With respect to the fixed points, it has been seen that there is a range of values of the mass that gives rise to non-trivial fixed points of the beta function, these fixed points are stable with respect to small perturbations around them, however, there is no trajectory flowing from the trivial fixed point to the non-trivial one because both are disconnected by a discontinuity of the beta function, in conclusion, the theory is not asymptotically safe.

We also see that the non-perturbative beta function that the method gives us in eq.(\ref{gblpa}) agrees with the perturbative results, successfully giving us the coefficients at one and two-loops of the beta function in the non-massive limit. 

In the next-to-leading order, we added the contributions of the momentum modes in the propagation of the chiral superfields and the vector superfield, to improve the calculation of the beta function by promoting the wave function renormalization functions to momentum-dependent functions through arbitrary functions in order to capture the influence of that modes in the beta function and study how they modify the NSVZ relation. In this case, a relation between the beta function and the anomalous dimension was again obtained in eq.(\ref{nlonsvz}), but the factor now depends on both the mass and the coefficients of the polynomial truncation, and it was shown that, given a fixed order of truncation, it is possible to define an effective fine-structure constant which has as its limit the result obtained in the local potential approximation when the coefficients are zero, and whose beta function satisfies the relation NSVZ, eq.(\ref{effcte}), resulting from the preservation of supersymmetry. A possible way to extend this study would be to analyze how these contributions modify the running of the fine-structure constant and, in the case of the next-to-leading order, to see if this beta function still agrees with the perturbative results.

The main contribution of this work has been to show that the results of the application of the functional renormalization group in supersymmetric gauge theories agree with the non-perturbative results obtained by geometric methods, and to provide a non-perturbative verification of the exact NSVZ beta function in the case of an abelian supersymmetric gauge theory; moreover, such results have been improved with the second order of the derivative expansion. Further investigations can explore the behavior of the modified beta function and the non-perturbative effects contributed by the momentum modes. These results can also be generalized to non-abelian supersymmetric gauge theories and more general superpotentials.

%% file: A/A.tex
Supersymmetric theories can be constructed from an action formulated in terms of its field content, as in eq.(\ref{sqed}), or in a more compacted form in terms of superfields over a superspace. Superfields are functions of both the space-time coordinates $\vt{x}{\gm}$, $\gm=(t,x,y,z)$ and the non-conmutative Grassmann variables $\lrt{\lsp{\go}{\ga},\rspc{\go}{\ga}}$. For $\mathcal{N}=1$ SQED we need only one pair of Grassmann variables.

A general superfield $\gS(x,\go,\bar\go)$ must be constrained to define chiral and vector superfields, which will be the supersymmetric version of matter and gauge superfields.

A chiral superfield $\gF(x,\go)$ is defined as such if it satisfies the conditions
\al{
\ft{\bar D}{\dot\ga}\gF=0\nn\\
\ft{D}{\ga}\hrm{\gF}=0
}{}
where
\al{
&\lsp{D}{\ga}=\pdif{}{\go^\ga}+i(\usm{\gm})_{\ga\dot\ga}\dv{\gm}\nn\\
&\rspc{D}{\ga}=-\pdif{}{\rsp{\go}{\ga}}-i\lspc{\go}{\ga}(\gs^\gm)_{\ga\dot\ga}\dv{\gm}.
}{}
are the covariant chiral/anti chiral derivatives.

The expansion of the chiral superfield in terms of Grassmann variables is given by
\al{
\gF(x,\go)=&\gf(x)+\sqrt{2}\go\gq(x)+\go\go F(x)+i\dv{\gm}\gf(x)\go\usm{\gm}\bar\go\nn\\
&-\frac{i}{\sqrt{2}}\go\go\dv{\gm}\gq(x)\usm{\gm}\bar\go-\inv{4}\dv{\gm}\idv{\gm}\gf(x)\go\go\bar\go\bar\go\nn\\
\hrm{\gF}(x,\bar\go)=&\hrm{\gf}(x)+\sqrt{2}\bar\go\bar\gq(x)+\bar\go\bar\go \hrm{F}(x)-i\dv{\gm}\hrm{\gf}(x)\go\usm{\gm}\bar\go\nn\\
&+\frac{i}{\sqrt{2}}\bar\go\bar\go\go\usm{\gm}\dv{\gm}\bar\gq(x)-\inv{4}\dv{\gm}\idv{\gm}\hrm{\gf}(x)\go\go\bar\go\bar\go.
}{}
To get a Lagrangian density in terms of the above superfield components, it is necessary to integrate the Grassmann coordinates. A renormalizable supersymmetric action for chiral superfields is
\al{
S=&\int d^4xd^2\go d^2\bar \go\gF^\dagger\gF=\int d^4x\dv{\gm}\hrm{\gf}\idv{\gm}\gf+i\bar\gq\bar\gs^\gm\dv{\gm}\gq+\hrm{F}F
}{}
This is known as the free Wess-Zumino model \cite{wess1974lagrangian}. The most general interaction for this model has the form \cite{aitchison2007supersymmetry}
\al{
\lag\stxt{int}=\pdif{W(\gf)}{\gf_i}F_i-\half\frac{\partial^2W(\gf)}{\partial\gf_i\partial\gf_j}\gq_i\cdot\gq_j+\textrm{h.c},
}{}
where the function $W(\gf)$ is the superpotential, an holomorphic function of the scalar fields. In superspace, the interaction action is
\al{
S\stxt{int}=\int d^4xd^2\go W(\gF)+\int d^4xd^2\bar\go \bar{W}(\hrm{\gF}).
}{}
On the other hand, $V(x,\go,\hrm{\go})$ is defined as a vector superfield if it satisfies the condition
\al{
V=V^\dagger.
}{}
This condition causes non-physical degrees of freedom, which can be gauge away using the transformation
\al{
V\to V+i\prt{\gW-\hrm{\gW}},
}{supergauge}
where $\gW$ is a chiral superfield. The transformation (\ref{supergauge}) is called a $U(1)$ supergauge transformation and we can semi-fix it so that it only holds for the usual gauge transformations. This is known as the Wess-Zumino gauge \cite{ferrara1974supergauge} and its expansion in Grassmann variables is
\al{
V\stxt{W-Z gauge}=\go\gs^\gm\bar\go\ft{A}{\gm}+i\go\go\bar\go\bar\gl-i\bar\go\bar\go\go\gl+\half \go\go\bar\go\bar\go D.
}{wzgauge}
Finally, $W_\ga$ is the gauge invariant abelian superfield.
\al{
\lsp{W}{\ga}\equiv \bar D^2\lsp{D}{\ga} V
}{}
which allows us to define a renormalizable supersymmetric and gauge invariant action for vector superfields.
\al{
\inv{32}\int d^4xd^2\go\vt{W}{\ga}\ft{W}{\ga}=\int d^4x-\inv{4}\ft{F}{\gm\gn}\vt{F}{\gm\gn}+i\bar\gl\bar\gs^\gm\dv{\gm}\gl+\half D^2.
}{}
In essence, supersymmetric gauge theories are the coupling between chiral and vector superfields. To describe charged fields, it is necessary to consider the left and right chiral components separately. Starting with two chiral superfields $\gF_1$ and $\gF_2$, we can define complex chiral superfields
\al{
\gF_-=&\frac{1}{\sqrt{2}}(\gF_1-i\gF_2)\nn\\
\gF_+=&\frac{1}{\sqrt{2}}(\gF_1+i\gF_2)\nn\\
}{chiralfields}
which, under the $U(1)$ gauge transformation, it transform as
\al{
\gF_\pm'\to\e{\pm 2ie\gW}\gF_\pm
}{}
then the products
\al{
&\hrm{\gF_{-}}\e{-2eV}\gF_{-},\nn\\
&\hrm{\gF_{+}}\e{2eV}\gF_{+}
}{}
are gauge invariant. Finally, the action for a supersymmetric $U(1)$ gauge theory will be
\al{
S=\int d^4xd^2\go d^2\bar \go\crt{\hrm{\gF_{-}}\e{-2eV}\gF_{-}+\hrm{\gF_{+}}\e{2eV}\gF_{+}}+\inv{32}\int d^4xd^2\go W^\ga W_\ga.
}{}
The action defined in eq.(\ref{sqedsuper}) has the vector superfield redefined as $V\to\dfrac{1}{e}V$ and a gauge-fixing term defined by the fact that in the Wess-Zumino gauge eq.(\ref{wzgauge})
\al{
{V\stxt{W-Z gauge}}^2(x,\go,\bar\go)=\half\go\go\bar\go\bar\go\ft{A}{\gm}\vt{A}{\gm}.
}{}
Finally, for the interaction part, we consider a superpotential with quadratic interaction
\al{
W(\gF_1,\gF_2)=&\half m\gF_1^2+\half m\gF_2^2+m\gF_1\gF_2+\text{h.c}.
}{}
Using the definitions in eq.(\ref{chiralfields}) the superpotential in terms of the electron and positron superfields is
\al{
W(\gF_-,\gF_+)=&m\gF_-\gF_++\text{h.c}.
}{}
Finally, we need a gauge-symmetry breaking term, analogous to the usual term used for the $R_\gj$ gauges. In supersymmetric gauge theories this term is \cite{ferrara1975perturbation}
\al{
\frac{2}{\gx e^2}\int d^4xd^2\go d^2\bar\go(D^2V)(\bar D^2V)=\inv{2\gx e^2}\dv{\gm}\vt{A}{\gm}\dv{\gn}\vt{A}{\gn},
}{}
from which we obtain the action in terms of the field content in eq.(\ref{sqed}). For more details of the superspace formalism see \cite{aitchison2007supersymmetry,martin2010supersymmetry,bailin1994supersymmetric}.

%% file: B/B.tex
The matrix representation of all operators used to calculate the Wetterich equation is given with respect to the following arrangement,
\al{
\gQ=&\lrt{\gf_-,\hrm{\gf_-},F_-,\hrm{F_-},\gf_+,\hrm{\gf_+},F_+,\hrm{F_+},D,A,\gq_-,\hrm{\bar\gq_-},\gq_+,\hrm{\bar{\gq}_{+}},\gl,\hrm{\bar\gl}},\nn\\
\hrm{\gQ}=&\lrt{\hrm{\gf_-},\gf_-,\hrm{F_-},F_-,\hrm{\gf_+},\gf_+,\hrm{F_+},F_+,D,A,\bar\gq_-,\hrm{\gq_-},\bar\gq_+,\hrm{\gq_+},\bar\gl,\hrm{\gl}}.
}{}
The convention for momentum space is
\al{
F(q)=\dfrac{1}{\prt{\sqrt{2\gp}}^d}\displaystyle\int d^dx F(x)\e{i\ft{q}{\gm}\vt{x}{\gm}}.
}{}
Dimensionless and renormalized electric charge is defined by
\al{
e_0=&Z_ek^{\frac{d-4}{2}}e,
}{}
similarly, the renormalized mass is\footnote{The fact that the mass is renormalized by the wave function renormalization of the chiral superfield is a consequence of the non-renormalization theorem.}
\al{
m_0=&Z_\gF km.
}{}
The form of dimensionless and renormalized regulators in eq.(\ref{reg}) are the optimized Litim regulator functions \cite{litim2000optimisation}
\al{
R\stxt{opt}(p)=&(k-p)\go(k^2-p^2)\nn\\
T\stxt{opt}(p^2)=&(k^2-p^2)\go(k^2-p^2)
}{litim}
defining dimensionless momentum
\al{
q=\frac{p}{k}
}{}
we can define eq.(\ref{litim}) in terms of a dimensionless regulator functions $r$, $t$
\al{
R\stxt{opt}(q)=&qr'(q),\nn\\
T\stxt{opt}(q^2)=&q^2t'(q)
}{}
and finally we choose $r'(q)$ and $t'(q^2)$ proportional to $Z_{\gF/V}$
\al{
r'(q)=Z_{\gF}r(q),\nn\\
t'(q)={Z_{V}}^2t(q^2).
}{}
In the calculation involving polynomial truncation of the momentum modes, the regulating functions that have been used are
\al{
r=&\gz_{\gF,k}(q)\prt{\frac{1}{q}\frac{\gz_{\gF,k}(1)}{\gz_{\gF,k}(q)}-1}\go\prt{1-q^2}\nn\\
t=&\gz_{\gF,k}(q)^2\prt{\frac{1}{q^2}\frac{\gz_{\gF,k}(1)^2}{\gz_{\gF,k}(q)^2}-1}\go\prt{1-q^2},
}{}
named as ``Litim-Type Regulator II'' in \cite{feldmann2016functional}. The advantage of these functions is that they eliminate dependencies on $q$ from the denominator for all orders of truncation.

%% file: C/C.tex
In eq.(\ref{wetterichlpa}) the flow of the LPA' truncation was computed using the operators
\al{
G_k=\gC_k^{(2)}+R_k
}{gilpa}
and
\al{
\dv{t}R_k.
}{drlpa}
The explicit forms of these operators are
\al{
G_k=\pmatx{G_{BB} & G_{BF}\\
G_{FB} & G_{FF}},\quad \dv{t}R_k=\pmatx{{\dv{t}R_k}_{BB} & 0\\
0 & {\dv{t}R_k}_{FF}}
}{}
\al{
&G_{BB}=\nn\\
&\pmatx{K_- & 0 & 0 & 0 & 0 & 0 & 0 & m_0 & Z_{\gF}\gf_- & -2Z_{\gF}A\gf_- \\
0 & K_- & 0 & 0 & 0 & 0 & m_0 & 0 & Z_{\gF}\hrm{\gf_-} & -2Z_{\gF}A\hrm{\gf_-}\\
0 & 0 & \mathcal{R} & 0 & 0 & m_0 & 0 & 0 & 0 & 0\\
0 & 0 & 0 & \mathcal{R} & m_0 & 0 & 0 & 0 & 0 & 0\\
0 & 0 & 0 & m_0 & K_+ & 0 & 0 & 0 & -Z_{\gF}\gf_+ & 2AZ_{\gF}\gf_+\\
0 & 0 & m_0 & 0 & 0 & K_+ & 0 & 0 & -Z_{\gF}\hrm{\gf_+} & 2AZ_{\gF}\hrm{\gf_+}\\
0 & m_0 & 0 & 0 & 0 & 0 & \mathcal{R} & 0 & 0 & 0\\
m_0 & 0 & 0 & 0 & 0 & 0 & 0 & \mathcal{R} & 0 & 0\\
Z_{\gF}\hrm{\gf_-} & Z_{\gF}\gf_- & 0 & 0 & -Z_{\gF}\hrm{\gf_+} & -Z_{\gF}\gf_+ & 0 & 0 & \mathcal{T} & 0\\
-2Z_{\gF}A\hrm{\gf_-} & -2Z_{\gF}A\gf_- & 0 & 0 & 2AZ_{\gF}\hrm{\gf_+} & 2AZ_{\gF}\gf_+ & 0 & 0 & 0 & J}
}{}
\al{
G_{BF}=\pmatx{0 & 0 & 0 & 0 & \sqrt{2}Z_\gF\gs_2\gq_-^T & 0 \\
0 & \sqrt{2}Z_\gF\gs_2\bar\gl & 0 & 0 & 0 & 0 \\
0 & 0 & 0 & 0 & 0 & 0 \\
0 & 0 & 0 & 0 & 0 & 0 \\
0 & 0 & 0 & 0 & -\sqrt{2}Z_\gF\gs_2\gq_+^T & 0 \\
0 & 0 & 0 & -\sqrt{2}Z_\gF\gs_2\bar\gl & 0 & 0 \\
0 & 0 & 0 & 0 & 0 & 0 \\
0 & 0 & 0 & 0 & 0 & 0 \\
0 & 0 & 0 & 0 & 0 & 0 \\
Z_\gF\bar\gs\bar\gq_- & 0 & Z_\gF\bar\gs\bar\gq_+ & 0 & 0 & 0}
}{}
\al{
G_{FB}=\pmatx{0 & 0 & 0 & 0 & 0 & 0 & 0 & 0 & 0 & Z_\gF\bar\gs\gq_- \\
0 & \sqrt{2}Z_\gF\gs_2\gl & 0 & 0 & 0 & 0 & 0 & 0 & 0 & 0 \\
0 & 0 & 0 & 0 & 0 & 0 & 0 & 0 & 0 & -Z_\gF\bar\gs\gq_+ \\
0 & 0 & 0 & 0 & 0 & -\sqrt{2}Z_\gF\gs_2\gl & 0 & 0 & 0 & 0 \\
\sqrt{2}Z_\gF\gs_2\bar\gq_-^T & 0 & 0 & 0 & -\sqrt{2}Z_\gF\gs_2\bar\gq_+^T & 0 & 0 & 0 & 0 & 0 \\
0 & 0 & 0 & 0 & 0 & 0 & 0 & 0 & 0 & 0 }
}{}
\al{
G_{FF}=\pmatx{\hat K_+ & 0 & 0 & 0 & 0 & 0 \\
0 & \mathcal{\hat R} & i\gs_2m & 0 & \sqrt{2}Z_\gF\gs_2\hrm{\gf_-} & 0 \\
0 & -i\gs_2m & \hat K_- & 0 & 0 & 0 \\
0 & 0 & 0 & \mathcal{\hat R} & -\sqrt{2}Z_\gF\gs_2\hrm{\gf_+} & 0 \\
0 & \sqrt{2}Z_\gF\gs_2\gf_- & 0 & -\sqrt{2}Z_\gF\gs_2\gf_+ & \mathcal{\hat T} & 0 \\
0 & 0 & 0 & 0 & 0 & \mathcal{\hat T}}
}{}

\al{
{\dv{t}R_k}_{BB}=\pmatx{q^2\dv{t}r & 0 & 0 & 0 & 0 & 0 & 0 & 0 & 0 & 0 \\
0 & q^2\dv{t}r & 0 & 0 & 0 & 0 & 0 & 0 & 0 & 0 \\
0 & 0 & \dv{t}r & 0 & 0 & 0 & 0 & 0 & 0 & 0 \\
0 & 0 & 0 & \dv{t}r & 0 & 0 & 0 & 0 & 0 & 0 \\
0 & 0 & 0 & 0 & q^2\dv{t}r & 0 & 0 & 0 & 0 & 0 \\
0 & 0 & 0 & 0 & 0 & q^2\dv{t}r & 0 & 0 & 0 & 0 \\
0 & 0 & 0 & 0 & 0 & 0 & \dv{t}r & 0 & 0 & 0 \\
0 & 0 & 0 & 0 & 0 & 0 & 0 & \dv{t}r & 0 & 0 \\
0 & 0 & 0 & 0 & 0 & 0 & 0 & 0 & \frac{1}{e^2}t& 0 \\
0 & 0 & 0 & 0 & 0 & 0 & 0 & 0 & 0 & \frac{q^2}{e^2\gx}t }
}{}
\al{
{\dv{t}R_k}_{FF}=\pmatx{\bar\gs\cdot q\dv{t}r & 0 & 0 & 0 & 0 & 0 \\
0 & \bar\gs\cdot q\dv{t}r & 0 & 0 & 0 & 0 \\
0 & 0 & \bar\gs\cdot q\dv{t}r & 0 & 0 & 0 \\
0 & 0 & 0 & \bar\gs\cdot q\dv{t}r & 0 & 0 \\
0 & 0 & 0 & 0 & \frac{\bar\gs\cdot q}{e^2}\dv{t}t & 0 \\
0 & 0 & 0 & 0 & 0 & \frac{\bar\gs\cdot q}{e^2}\dv{t}t }
}{}
where
\al{
&\mathcal{R}=r+Z_{\gF}\nn\\
&\mathcal{T}=\inv{e^2}\prt{t+{Z_V}^2}\nn\\
&\mathcal{\hat R}=\bar\gs\cdot q\mathcal{R}\nn\\
&\mathcal{\hat T}=\bar\gs\cdot q\mathcal{R}\nn\\
&K_\mp=q^2\mathcal{R}\mp Z_\gF D\pm Z_\gF A^2\\
&J=\frac{q^2}{\gx}\mathcal{T}-2Z_\gF\gf_-\hrm{\gf_-}+2Z_\gF\gf_+\hrm{\gf_+}\nn\\
&\hat K_\pm=\mathcal{\hat R}\pm Z_\gF\bar\gs\cdot A.
}{}
Making the product between (\ref{gilpa}) and (\ref{drlpa}) and taking the trace, we are able to perform eq.(\ref{betaeqlpa}) and eq.(\ref{anomalpa}). Using the dimensionless and renormalized quantities of Appendix \ref{sec:B} and performing eq.(\ref{anomalpa}) we obtain the following integral in $d-$dimensions for the anomalous dimension of the chiral superfield
\al{
\gc=&\frac{\gW_d}{(2\gp)^d}\int\ud{0}{1} dqq^{d-1}\frac{e_R^2\crt{2\gr(q)\gt(q)\mathcal{M}_+(q)\dv{t}r(q)-\gr(q)^2\mathcal{M}_-(q)\dv{t}t(q)}}{\mathcal{M}_-(q)^3\gt(q)^2}\nn\\
=&\frac{\gW_d}{(2\gp)^d}\int\ud{0}{1} dqq^{d-1}\frac{e_R^2\crt{2\gr(q)\gt(q)\mathcal{M}_+(q)(-(1-d)\gc-q\dv{q})r(q)-\gr(q)^2\mathcal{M}_-(q)(4\gh_V-q\dv{q})t(q)}}{\mathcal{M}_-(q)^3\gt(q)^2}
}{anomalousdlpa}
and for the anomalous dimension of the vector superfield
\al{
-\frac{1}{2}\gh_V=&\frac{\gW_d}{(2\gp)^d}\int\ud{0}{1} dqq^{d-1}\frac{e_R^2\gr(q)\mathcal{M}_+(q)\dv{t}r(q)}{\mathcal{M}_-(q)^3}\nn\\
=&\frac{\gW_d}{(2\gp)^d}\int\ud{0}{1} dqq^{d-1}\frac{e_R^2\gr(q)\mathcal{M}_+(q)(-(1-d)\gc-q\dv{q})r(q)}{\mathcal{M}_-(q)^3},
}{betadlpa}
where $\gW_d=\dfrac{2(\gp)^{\frac{d}{2}}}{\gC\crt{\frac{d}{2}}}$ is the surface of a $d-1$ dimensional sphere and
\al{
&\gr(q)=r(q)+1\nn\\
&\gt(q)=t(q)+1\nn\\
&\mathcal{M}_\pm(q)=m_0^2\pm q^2\gr(q)^2.
}{}
Replacing eq.(\ref{reg}) we can solve the integrals of eq.(\ref{anomalousdlpa}) and eq.(\ref{betadlpa}) in $d=4$ dimensions, which gives us eq.(\ref{lpagamma}) and eq.(\ref{lpabeta}).

It is also necessary to add that it is possible to find the result of eq(\ref{betadlpa}) using a different projection, if we set $\gf_-=\gf_+=\gq_-=\gq_+=F_+=F_-=A=D=0$ we get
\al{
\dv{t}\gC_k=(\sqrt{2\gp})^4\dd{0}{4}\dv{t}\crt{\frac{{Z_{V}}^2}{e^2}i\usmb{\gm}\ft{k}{\gm}\bar\gl\gl},
}{}
therefore eq.(\ref{betadlpa}) can be found from 
\al{
-\frac{1}{2}\gh_V=\prt{\frac{e^2}{2{Z_V}^2}\artr{\lrt{\artr{\frac{\D^2}{\D\bar\gl\D\gl}\dv{t}\gC_k}_{\gf_-=\gf_+=\gq_-=\gq_+=F_+=F_-=A=D=0}}}_{\gl=0}}_{i\usmb{\gm}\ft{k}{\gm}\text{-coeff}}, 
}{}
emphasizing that it is the coefficient of the expression that accompanies $i\usmb{\gm}\ft{k}{\gm}$.

%% file: D/D.tex
In eq.(\ref{desc}) we decompose the operator $G_k$ into $G_{0,k}$ and $\gD G_k$. The explicit forms of the operators are
\al{
G_{0,k}=\pmatx{{G_{0,k}}_{BB} & 0\\
0 & {G_{0,k}}_{FF}}
}{}
\al{
{G_{0,k}}_{BB}=\pmatx{q^2\mathcal{R} & 0 & 0 & 0 & 0 & 0 & 0 & m_0 & 0 & 0 \\
0 & q^2\mathcal{R} & 0 & 0 & 0 & 0 & m_0 & 0 & 0 & 0\\
0 & 0 & \mathcal{R} & 0 & 0 & m_0 & 0 & 0 & 0 & 0\\
0 & 0 & 0 & \mathcal{R} & m_0 & 0 & 0 & 0 & 0 & 0\\
0 & 0 & 0 & m_0 & q^2\mathcal{R} & 0 & 0 & 0 & 0 & 0\\
0 & 0 & m_0 & 0 & 0 & q^2\mathcal{R} & 0 & 0 & 0 & 0\\
0 & m_0 & 0 & 0 & 0 & 0 & \mathcal{R} & 0 & 0 & 0\\
m_0 & 0 & 0 & 0 & 0 & 0 & 0 & \mathcal{R} & 0 & 0\\
0 & 0 & 0 & 0 & 0 & 0 & 0 & 0 & \mathcal{T} & 0\\
0 & 0 & 0 & 0 & 0 & 0 & 0 & 0 & 0 & \inv{\gx}\mathcal{T}}
}{}
\al{
{G_{0,k}}_{FF}=\pmatx{\mathcal{\hat R} & 0 & 0 & 0 & 0 & 0 & 0 & 0 & 0 & 0 \\
0 & \mathcal{\hat R} & i\frac{m_0}{2}\gs_2 & 0 & 0 & 0 & 0 & 0 & 0 & 0\\
0 & -i\frac{m_0}{2}\gs_2 & \mathcal{\hat R} & 0 & 0 & 0 & 0 & 0 & 0 & 0\\
0 & 0 & 0 & \mathcal{\hat R} & 0 & 0 & 0 & 0 & 0 & 0\\
0 & 0 & 0 & 0 & \mathcal{\hat R} & 0 & 0 & 0 & 0 & 0\\
0 & 0 & 0 & 0 & 0 & \mathcal{\hat R} & 0 & 0 & 0 & 0\\
0 & 0 & 0 & 0 & 0 & 0 & \mathcal{\hat R} & 0 & 0 & 0\\
0 & 0 & 0 & 0 & 0 & 0 & 0 & \mathcal{\hat R} & 0 & 0\\
0 & 0 & 0 & 0 & 0 & 0 & 0 & 0 & \mathcal{\hat T} & 0\\
0 & 0 & 0 & 0 & 0 & 0 & 0 & 0 & 0 & \mathcal{\hat R}}
}{}
\al{
\gD G_k=\pmatx{{\gD G_k}_{BB} & {\gD G_k}_{BF}\\
{\gD G_k}_{FB} & {\gD G_k}_{FF}}
}{}
\al{
&{\gD G_k}_{BB}=\nn\\
&\pmatx{K_- & 0 & 0 & 0 & 0 & 0 & 0 & 0 & Z_{\gF}\gf_- & -2Z_{\gF}A\gf_- \\
0 & K_- & 0 & 0 & 0 & 0 & 0 & 0 & Z_{\gF}\hrm{\gf_-} & -2Z_{\gF}A\hrm{\gf_-}\\
0 & 0 & \mathcal{R} & 0 & 0 & 0 & 0 & 0 & 0 & 0\\
0 & 0 & 0 & \mathcal{R} & 0 & 0 & 0 & 0 & 0 & 0\\
0 & 0 & 0 & 0 & K_+ & 0 & 0 & 0 & -Z_{\gF}\gf_+ & 2AZ_{\gF}\gf_+\\
0 & 0 & 0 & 0 & 0 & K_+ & 0 & 0 & -Z_{\gF}\hrm{\gf_+} & 2AZ_{\gF}\hrm{\gf_+}\\
0 & 0 & 0 & 0 & 0 & 0 & \mathcal{R} & 0 & 0 & 0\\
0 & 0 & 0 & 0 & 0 & 0 & 0 & \mathcal{R} & 0 & 0\\
Z_{\gF}\hrm{\gf_-} & Z_{\gF}\gf_- & 0 & 0 & -Z_{\gF}\hrm{\gf_+} & -Z_{\gF}\gf_+ & 0 & 0 & \mathcal{T} & 0\\
-2Z_{\gF}A\hrm{\gf_-} & -2Z_{\gF}A\gf_- & 0 & 0 & 2AZ_{\gF}\hrm{\gf_+} & 2AZ_{\gF}\gf_+ & 0 & 0 & 0 & J}
}{}
\al{
{\gD G_k}_{BF}=\pmatx{0 & 0 & 0 & 0 & \sqrt{2}Z_\gF\gs_2\gq_-^T & 0 \\
0 & \sqrt{2}Z_\gF\gs_2\bar\gl & 0 & 0 & 0 & 0 \\
0 & 0 & 0 & 0 & 0 & 0 \\
0 & 0 & 0 & 0 & 0 & 0 \\
0 & 0 & 0 & 0 & -\sqrt{2}Z_\gF\gs_2\gq_+^T & 0 \\
0 & 0 & 0 & -\sqrt{2}Z_\gF\gs_2\bar\gl & 0 & 0 \\
0 & 0 & 0 & 0 & 0 & 0 \\
0 & 0 & 0 & 0 & 0 & 0 \\
0 & 0 & 0 & 0 & 0 & 0 \\
Z_\gF\bar\gs\bar\gq_- & 0 & Z_\gF\bar\gs\bar\gq_+ & 0 & 0 & 0}
}{}
\al{
&{\gD G_k}_{FB}=\pmatx{0 & 0 & 0 & 0 & 0 & 0 & 0 & 0 & 0 & Z_\gF\bar\gs\gq_- \\
0 & \sqrt{2}Z_\gF\gs_2\gl & 0 & 0 & 0 & 0 & 0 & 0 & 0 & 0 \\
0 & 0 & 0 & 0 & 0 & 0 & 0 & 0 & 0 & -Z_\gF\bar\gs\gq_+ \\
0 & 0 & 0 & 0 & 0 & -\sqrt{2}Z_\gF\gs_2\gl & 0 & 0 & 0 & 0 \\
\sqrt{2}Z_\gF\gs_2\bar\gq_-^T & 0 & 0 & 0 & -\sqrt{2}Z_\gF\gs_2\bar\gq_+^T & 0 & 0 & 0 & 0 & 0 \\
0 & 0 & 0 & 0 & 0 & 0 & 0 & 0 & 0 & 0 }
}{}
\al{
{\gD G_k}_{FF}=\pmatx{\hat K_+ & 0 & 0 & 0 & 0 & 0 \\
0 & \mathcal{\hat R} & 0 & 0 & \sqrt{2}Z_\gF\gs_2\hrm{\gf_-} & 0 \\
0 & 0 & \hat K_- & 0 & 0 & 0 \\
0 & 0 & 0 & \mathcal{\hat R} & -\sqrt{2}Z_\gF\gs_2\hrm{\gf_+} & 0 \\
0 & \sqrt{2}Z_\gF\gs_2\gf_- & 0 & -\sqrt{2}Z_\gF\gs_2\gf_+ & \mathcal{\hat T} & 0 \\
0 & 0 & 0 & 0 & 0 & \mathcal{\hat T}}
}{}
\al{
\dv{t}R_k=\pmatx{{\dv{t}R_k}_{BB} & 0\\
0 & {\dv{t}R_k}_{FF}}
}{}
\al{
{\dv{t}R_k}_{BB}=\pmatx{q^2\dv{t}r & 0 & 0 & 0 & 0 & 0 & 0 & 0 & 0 & 0 \\
0 & q^2\dv{t}r & 0 & 0 & 0 & 0 & 0 & 0 & 0 & 0 \\
0 & 0 & \dv{t}r & 0 & 0 & 0 & 0 & 0 & 0 & 0 \\
0 & 0 & 0 & \dv{t}r & 0 & 0 & 0 & 0 & 0 & 0 \\
0 & 0 & 0 & 0 & q^2\dv{t}r & 0 & 0 & 0 & 0 & 0 \\
0 & 0 & 0 & 0 & 0 & q^2\dv{t}r & 0 & 0 & 0 & 0 \\
0 & 0 & 0 & 0 & 0 & 0 & \dv{t}r & 0 & 0 & 0 \\
0 & 0 & 0 & 0 & 0 & 0 & 0 & \dv{t}r & 0 & 0 \\
0 & 0 & 0 & 0 & 0 & 0 & 0 & 0 & \frac{1}{e^2}t& 0 \\
0 & 0 & 0 & 0 & 0 & 0 & 0 & 0 & 0 & \frac{q^2}{e^2\gx}t }
}{}
\al{
{\dv{t}R_k}_{FF}=\pmatx{\bar\gs\cdot q\dv{t}r & 0 & 0 & 0 & 0 & 0 \\
0 & \bar\gs\cdot q\dv{t}r & 0 & 0 & 0 & 0 \\
0 & 0 & \bar\gs\cdot q\dv{t}r & 0 & 0 & 0 \\
0 & 0 & 0 & \bar\gs\cdot q\dv{t}r & 0 & 0 \\
0 & 0 & 0 & 0 & \frac{\bar\gs\cdot q}{e^2}\dv{t}t & 0 \\
0 & 0 & 0 & 0 & 0 & \frac{\bar\gs\cdot q}{e^2}\dv{t}t }
}{}
where
\al{
&\mathcal{R}=r+Z_{\gF,k}\nn\\
&\mathcal{T}=\inv{e^2}\prt{t+{Z_{V,k}}^2}\nn\\
&\mathcal{\hat R}=\bar\gs\cdot q\mathcal{R}\nn\\
&\mathcal{\hat T}=\bar\gs\cdot q\mathcal{R}\nn\\
&K_\mp=\mp Z_{\gF,k} D\pm Z_{\gF,k} A^2\\
&J=-2Z_{\gF,k}\gf_-\hrm{\gf_-}+2Z_{\gF,k}\gf_+\hrm{\gf_+}\nn\\
&\hat K_\pm=\pm Z_{\gF,k}\bar\gs\cdot A.
}{}
We need the quadratic term in the field-dependent part of eq.(\ref{wetterichnloex}). Defining
\al{
W(q,q')=G_{0,k}^{-1}(q)\gD G_{k}(q+q')
}{}
we can see that the quadratic terms have the form
\al{
W^2(q,q')=\int d^4q''W(q,q'')W(q'',q')=\int d^4q''G_{0,k}^{-1}(q)\gD G_{k}(q''+q)G_{0,k}^{-1}(q'')\gD G_{k}(q''+q'),
}{}
whose trace corresponds to integrating in $q$ by doing $q'\to q$. Thus
\al{
-\inv{4}\text{STr}\wdv{t}&\prt{G_{0,k}^{-1}\gD G_{k}}^2=\nn\\
&-\inv{4}\int d^4q d^4q'G_{0,k}^{-1}(q)\dv{t}R_k(q)G_{0,k}^{-1}(q)\gD G_{k}(q'+q)G_{0,k}^{-1}(q')\gD G_{k}(q'+q)\nn\\
&+\inv{4}\int d^4q d^4q'G_{0,k}^{-1}(q)\gD G_{k}(q'+q)G_{0,k}^{-1}(q')\dv{t}R_k(q')G_{0,k}^{-1}(q')\gD G_{k}(q'+q).
}{}
similarly for the cubic term,
\al{
W^3&(q,q')=\int d^4q''d^4q'''W(q,q'')W(q'',q''')W(q''',q')\nn\\
&=\int d^4q''d^4q'''G_{0,k}^{-1}(q)\gD G_{k}(q''+q)G_{0,k}^{-1}(q'')\gD G_{k}(q''+q''')G_{0,k}^{-1}(q''')\gD G_{k}(q'''+q'),
}{}
the trace corresponds to integrating in $q$ by doing $q'\to q$. Thus
\al{
&\inv{6}\text{STr}\wdv{t}\prt{G_{0,k}^{-1}\gD G_{k}}^3=\nn\\
&\inv{6}\int d^4q d^4q'd^4q''G_{0,k}^{-1}(q)\dv{t}R_k(q)G_{0,k}^{-1}(q)\gD G_{k}(q'+q)G_{0,k}^{-1}(q')\gD G_{k}(q'+q'')G_{0,k}^{-1}(q'')\gD G_{k}(q''+q)\nn\\
&+\inv{6}\int d^4q d^4q'd^4q''G_{0,k}^{-1}(q)\gD G_{k}(q'+q)G_{0,k}^{-1}(q')\dv{t}R_k(q')G_{0,k}^{-1}(q')\gD G_{k}(q'+q'')G_{0,k}^{-1}(q'')\gD G_{k}(q''+q)\nn\\
&+\inv{6}\int d^4q d^4q'd^4q''G_{0,k}^{-1}(q)\gD G_{k}(q'+q)G_{0,k}^{-1}(q')\gD G_{k}(q'+q'')G_{0,k}^{-1}(q'')\dv{t}R_k(q'')G_{0,k}^{-1}(q'')\gD G_{k}(q''+q).
}{}
By performing the matrix products, taking the matrix trace, and computing the $q',q''-$integrals, we obtain the expressions in eq.(\ref{nloint}) after performing the corresponding functional derivatives.

%% file: E/E.tex
The anomalous dimension and beta function in terms of the effective fine-structure constant of eq.(\ref{effgbm}) are
\al{
\td\gc(\td\ga)=&\frac{\bar{g}'_{0,2}\td\ga^2+\bar{g}'_{0,1}\td\ga}{\bar{g}'_{1,2}\td\ga^2+\bar{g}'_{1,1}\td\ga+\bar{g}'_{1,0}}
}{}
and
\al{
\td\gb(\td\ga)=&\frac{\bar{b}_{0,3}\td\ga^3+\bar{b}_{0,2}\td\ga^2}{\bar{b}_{1,2}\td\ga^2+\bar{b}_{1,1}\td\ga+\bar{b}_{1,0}}
}{}
where
\al{
\bar{g}_{0,2}=&5 (-14 - 3 \gz_{V,1} (12 + 7 \gz_{V,1})) (1 + \gz_{\gF,1}) (-m^2 + (1 +\gz_{\gF,1})^2)\nn\\
\bar{g}_{0,1}=&5 \gp 84 + 28 \gz_{\gF,1} \crt{10 + m^2 + \gz_{\gF,1} (11 + 4 \gz_{\gF,1})} + 7 \gz_{V,1}^2 \crtr{m^2 (-3 + \gz_{\gF,1})}\nn\\
&\lrtl{\crtr{+ (1 + \gz_{\gF,1})^2
(15 + 19 \gz_{\gF,1})}}\lrtr{+ 8 \gz_{V,1} \crt{m^2 (-3 + 4 \gz_{\gF,1}) + (1 + \gz_{\gF,1})^2 (24 + 31 \gz_{\gF,1})}}\nn\\
\bar{g}_{1,2}=&-5 \crt{14 + 3 \gz_{V,1} (12 + 7 \gz_{V,1})} (1 + \gz_{\gF,1}) \prt{-m^2 + (1 +
\gz_{\gF,1})^2}\nn\\
\bar{g}_{1,1}=&42 \gp (1 + \gz_{V,1})^2 (3 + 5 \gz_{\gF,1}) \prt{m^2 + (1 + \gz_{\gF,1})^2}\nn\\
\bar{g}_{1,1}=&-420 \gp^2 (1 + \gz_{V,1})^4 (2 + 3 \gz_{\gF,1}) \prt{m^2 + (1 + \gz_{\gF,1})^2}\nn\\
\bar{b}_{0,3}=&m^2 \crt{126 + 3 \gz_{V,1} (124 + 77 \gz_{V,1}) + 70 \gz_{\gF,1} + 5 \gz_{V,1} (52 + 35 \gz_{V,1}) \gz_{\gF,1}} \nn\\
&- (1 + \gz_{\gF,1})^2 \lrt{294 + 350 \gz_{\gF,1} + \gz_{V,1} \crt{708 + 820 \gz_{\gF,1} + 7 \gz_{V,1} (57 + 65 \gz_{\gF,1})}}\nn\\
\bar{b}_{0,2}=&-420 \gp (1 + \gz_{V,1})^4 (2 + 3 \gz_{\gF,1}) \prt{m^2 + (1 + \gz_{\gF,1})^2}\nn\\
\bar{b}_{1,2}=&10 \crt{14 + 3 \gz_{V,1} (12 + 7 \gz_{V,1})} (1 + \gz_{\gF,1}) (m^2 - (1 +
\gz_{\gF,1})^2)\nn\\
\bar{b}_{1,1}=&84 \gp (1 + \gz_{V,1})^2 (3 + 5 \gz_{\gF,1}) \prt{m^2 + (1 + \gz_{\gF,1})^2}\nn\\
\bar{b}_{1,0}=&-420 \gp^2 (1 + \gz_{V,1})^4 (4 + 6 \gz_{\gF,1}) \prt{m^2 + (1 + \gz_{\gF,1})^2}.
}{}